\documentclass[rnote, traditabstract]{aa}

\usepackage{natbib}
\usepackage{graphicx}
\usepackage{txfonts}

\begin{document}

\title{On the Modified Random Walk algorithm for Monte-Carlo Radiation Transfer}
\author{T. P. Robitaille\inst{1}}
\institute{Spitzer Postdoctoral Fellow, Harvard-Smithsonian Center for Astrophysics, 60 Garden Street, Cambridge, MA, 02138, USA\\
\email{trobitaille@cfa.harvard.edu}}

\date{Received 20 May 2010 / Accepted 12 September 2010}
 
\abstract{Min et al. (2009) presented two complementary techniques that use the diffusion approximation to allow efficient Monte-Carlo radiation transfer in very optically thick regions: a modified random walk and a partial diffusion approximation. In this note, I show that the calculations required for the modified random walk method can be significantly simplified. In particular, the diffusion coefficient and the mass absorption coefficients required for the modified random walk are in fact the same as the standard diffusion coefficient and the Planck mean mass absorption coefficient.}{}


\titlerunning{On the Modified Random Walk for Monte-Carlo Radiation Transfer}
\authorrunning{T. P. Robitaille}

\maketitle

\section{Introduction}

   The problem of Monte-Carlo radiation transfer in very optically thick regions -- such as in the midplane of circumstellar disks -- is challenging. Without any approximations, photon packets can get trapped for millions of interactions, increasing the required computational time by several orders of magnitude. \citet[][hereafter M09]{Min:09:155} presented two complementary methods to greatly improve the efficiency of Monte-Carlo radiation transfer codes in very optically thick regions: a modified random walk (MRW) and a partial diffusion approximation (PDA). The MRW prevents photons from getting stuck in very optically thick regions, and the PDA allows temperatures to be calculated in regions that see few or no photons.
   
   The essence of the MRW method is that instead of computing thousands to millions of individual absorption or scattering events for a single photon in these optically thick regions, one can make use of the solution to the diffusion approximation inside small regions to propagate the photon efficiently. Monte-Carlo radiation transfer codes propagate photons in a grid made up of cells of constant density and temperature. Therefore if the mean optical depth to the edge of a cell is much larger than unity, one can set up a sphere whose radius is smaller than the distance to the closest wall, inside which the density will be constant, and travel to the edge of a sphere in a single step using the diffusion approximation.
   
   A probability distribution function is used to sample the true distance traveled to exit this sphere (since the photon would follow a random walk inside the sphere, rather than moving in a straight line). This true distance, which depends on the radius of the sphere and the local diffusion coefficient $D$, can then be used along with the mass absorption coefficient $\bar{\kappa}$ to compute the total amount of energy deposited in the dust during the diffusion. This is required in order to compute the temperature in the cell accurately. Finally, the photon is emitted from a random position on the surface of the sphere with a frequency sampled from the Planck function.
   
   M09 provide equations for $D$, $\bar{\kappa}$, and the dust emission coefficient $\eta_\nu$, taking into account that photons can be both scattered and absorbed and re-emitted. In M09, the suggested algorithm is to first calculate $\eta_\nu$ iteratively, and to then use $\eta_\nu$ to compute $D$ and $\bar{\kappa}$. In this note, I show that $\eta_\nu$ does not need to be solved iteratively, but can be solved directly, and I use this solution to show that $D$ and $\bar{\kappa}$ can in fact very easily be computed, resulting in both a simpler implementation of the MRW, and in some cases performance gains.

\section{Derivation}

\label{sec:derivation}

\subsection{Emission coefficient}

\label{sec:correct}

In the presence of isotropic scattering, the emissivity of dust in local thermodynamic equilibrium (LTE) is given by
\begin{equation}
\label{eq:initial}
\eta_\nu = \kappa_\nu\,B_\nu(T)+ \sigma_\nu\,J_\nu
\end{equation}
where $\nu$ is the frequency of the radiation,  $\kappa_\nu$ is the mass absorption coefficient, $B_\nu(T)$ is the Planck function at the temperature $T$ of the dust, $\sigma_\nu$ is the mass scattering coefficient, and $J_\nu$ is the mean intensity of the radiation field. Assuming that the radiation is isotropic, $J_\nu = I_\nu$, where $I_\nu$ is the intensity of the radiation. In the  optically thick regime, $I_\nu=S_\nu$, where $S_\nu$ is the source function. Therefore,
\begin{equation}
\label{eq:initials}
\eta_\nu = \kappa_\nu\,B_\nu(T) + \sigma_\nu\,S_\nu
\end{equation}
The source function is defined as the ratio of the total emissivity to the total extinction, which in this case is
\begin{equation}
S_\nu = \frac{\eta_\nu}{\chi_\nu},
\end{equation}
where $\chi_\nu$ is the mass extinction coefficient ($\chi_\nu = \kappa_\nu + \sigma_\nu$). Therefore, Equation (\ref{eq:initials}) can be rewritten as
\begin{equation}
\label{eq:correct}
\eta_\nu = \kappa_\nu\,B_\nu(T) + \sigma_\nu\,\frac{\eta_\nu}{\chi_\nu}.
\end{equation}
Re-arranging this equation, one obtains
\begin{equation}
\eta_\nu = \frac{\kappa_\nu\,B_\nu(T)}{1 - \sigma_\nu/\chi_\nu},
\end{equation}
which can be simplified, since:
\begin{equation}
\frac{\kappa_\nu}{1-\sigma_\nu/\chi_\nu} = \frac{\kappa_\nu}{1-(\chi_\nu-\kappa_\nu)/\chi_\nu} = \frac{\kappa_\nu}{\kappa_\nu/\chi_\nu} = \chi_\nu.
\end{equation}
Therefore,
\begin{equation}
\label{eq:finaleta}
\eta_\nu = \chi_\nu\,B_\nu(T)
\end{equation}
The source function for this emissivity is
\begin{equation}
S_\nu = \frac{\eta_\nu}{\chi_\nu} = B_\nu(T),
\end{equation}
which is expected for thermal emission from dust in LTE in the optically thick regime.

\subsection{Comparison with the M09 emission coefficient}

In Equation (13) of their paper, M09 wrote the emission coefficient for dust in an optically thick region as:
\begin{equation}
\label{eq:etamin}
\eta_\nu = \kappa_\nu\,B_\nu(T) \,\frac{\displaystyle{\int_0^\infty\eta_\nu\,\frac{\kappa_\nu}{\chi_\nu}\,d\nu}}{\displaystyle{\int_0^\infty\eta_\nu\,d\nu}}+ \frac{\sigma_\nu}{\chi_\nu}\eta_\nu.
\end{equation}
Their original equation included a $dB_\nu(T)/dT$ term instead of $B_\nu(T)$, because they make use of the \citet{Bjorkman:01:615} temperature correction method for determining the dust temperature. I use $B_\nu(T)$ here for consistency, but throughout the derivation,  $B_\nu(T)$ can be replaced by $dB_\nu(T)/dT$ with no impact on the final result.

Equation (\ref{eq:etamin}) is similar to Equation (\ref{eq:correct}), but includes an extra term which is the ratio of two integrals. It is not clear why this term was included by M09, because other than possibly changing the dust temperature -- which would affect $B_\nu(T)$ -- the addition of scattering should not affect the thermal emissivity if $\kappa_\nu$ is held constant. However, even with this extra term, Equation (\ref{eq:etamin}) can in fact be simplified.
One can set 
\begin{equation}
\label{eq:zeta}
\epsilon = \frac{\displaystyle{\int_0^\infty\eta_\nu\,\frac{\kappa_\nu}{\chi_\nu}\,d\nu}}{\displaystyle{\int_0^\infty\eta_\nu\,d\nu}}
\end{equation}
which is a frequency-independent constant for any given dust type. Equation (\ref{eq:etamin}) then becomes:
\begin{equation}
\eta_\nu = \kappa_\nu\,B_\nu(T)\,\epsilon+\frac{\sigma_\nu}{\chi_\nu}\,\eta_\nu
\end{equation}
M09 suggest solving this through an iterative method, but in fact, this can be solved exactly, by re-arranging for $\eta_\nu$ as for Equation (\ref{eq:correct}) in \S\ref{sec:correct},  giving:
\begin{equation}
\eta_\nu = \chi_\nu\,B_\nu(T)\,\epsilon
\end{equation}
Substituting this back into Equation (\ref{eq:zeta}) gives
\begin{equation}
\epsilon = \frac{\displaystyle{\int_0^\infty\chi_\nu\,B_\nu(T)\,\epsilon\,\frac{\kappa_\nu}{\chi_\nu}\,d\nu}}{\displaystyle{\int_0^\infty\chi_\nu\,B_\nu(T)\,\epsilon\,\,d\nu}}
\end{equation}
The $\epsilon$ in the integrals cancel out (since they are frequency independent), so that
\begin{equation}
\epsilon = \frac{\displaystyle{\int_0^\infty\kappa_\nu\,B_\nu(T)\,d\nu}}{\displaystyle{\int_0^\infty\chi_\nu\,B_\nu(T)\,\,d\nu}} = \frac{\bar{\kappa}_P}{\bar{\chi}_P},
\end{equation}
where $\bar{\kappa}_P$ and $\bar{\chi}_P$ are the Planck mean mass absorption and extinction coefficients respectively. Therefore, Equation (\ref{eq:etamin}) can solved exactly rather than through an interative procedure, with the solution given by
\begin{equation}
\label{eq:finaleta_complete}
\eta_\nu = \chi_\nu\,B_\nu(T)\,\frac{\bar{\kappa}_P}{\bar{\chi}_P}
\end{equation}
This is very similar to Equation (\ref{eq:finaleta}), but includes an extra constant multiplicative term. However, the derivations in the following sections are valid regardless of whether the emissivity is calculated using Equation (\ref{eq:finaleta}) or (\ref{eq:finaleta_complete}), as in all cases multiplicative constants to the emissivity cancel out.

\subsection{Diffusion coefficient}

The diffusion coefficient is given by M09 as
\begin{equation}
D = \frac{\langle d^2\rangle}{6\,\langle d\rangle} = \frac{1}{3}\, \frac{\displaystyle{\int_0^\infty\frac{\eta_\nu}{\rho^2\chi^2_\nu}\,d\nu}}{\displaystyle{\int_0^\infty\frac{\eta_\nu}{\rho\chi_\nu}\,d\nu}}
\end{equation}
where $\langle d\rangle$ is the mean free path of the photons, $\langle d^2\rangle$ is the mean of the path lengths squared, and $\rho$ is the density of the dust. Substituting either Equation  (\ref{eq:finaleta}) or (\ref{eq:finaleta_complete}) into the above gives
\begin{equation}
D = \frac{1}{3}\, \frac{\displaystyle{\int_0^\infty\frac{\chi_\nu B_\nu(T)}{\rho^2\chi^2_\nu}\,d\nu}}{\displaystyle{\int_0^\infty\frac{\chi_\nu B_\nu(T)}{\rho\chi_\nu}\,d\nu}}
\end{equation}
The $\chi_\nu$ terms can be simplified, and the $\rho$ term can factored out as it is not dependent on frequency:
\begin{equation}
D = \frac{1}{3\rho}\, \frac{\displaystyle{\int_0^\infty\frac{B_\nu(T) }{\chi_\nu}\,d\nu}}{\displaystyle{\int_0^\infty B_\nu(T)\,d\nu}}
\end{equation}
The second term can easily be recognized as $1/\bar{\chi}_R$, the inverse of the Rosseland mean mass extinction coefficient. Thus,
\begin{equation}
\label{eq:diffcoeff}
D = \frac{1}{3\rho\bar{\chi}_R}
\end{equation}
This means that the expression for the diffusion coefficient is in fact the same whether or not scattering is included.

\subsection{Mean opacity to absorption}

The mean opacity to absorption in the diffusion region is 
\begin{equation}
\label{eq:kappabar}
\bar{\kappa} = \frac{\displaystyle{\int_0^\infty\frac{\eta_\nu }{\chi_\nu}\,\kappa_\nu\,d\nu}}{\displaystyle{\int_0^\infty \frac{\eta_\nu }{\chi_\nu}\,d\nu}}.
\end{equation}
This is a mean frequency-dependent mass absorption coefficient weighted by the probability of emission at a given frequency $\eta_\nu$, and by the mean free path at each frequency, $1/\rho\chi_\nu$ (where the $\rho$ term cancels out). Substituting Equation (\ref{eq:finaleta}) or (\ref{eq:finaleta_complete}) into Equation (\ref{eq:kappabar}) gives
\begin{equation}
\bar{\kappa} = \frac{\displaystyle{\int_0^\infty\frac{\chi_\nu\,B_\nu(T)}{\chi_\nu}\,\kappa_\nu\,d\nu}}{\displaystyle{\int_0^\infty \frac{\chi_\nu\,B_\nu(T)}{\chi_\nu}\,d\nu}}.
\end{equation}
As before, this can be simplified by canceling the $\chi_\nu$ terms:
\begin{equation}
\bar{\kappa} = \frac{\displaystyle{\int_0^\infty B_\nu(T)\,\kappa_\nu\,d\nu}}{\displaystyle{\int_0^\infty B_\nu(T)\,d\nu}} = \bar{\kappa}_P,
\end{equation}
which is the standard Planck mean mass absorption coefficient.

\section{Implementation}

\label{sec:implementation}

In this section, I summarize the M09 MRW algorithm, in the light of the new equations derived in \S\ref{sec:derivation}. A good criterion for deciding whether to start the MRW procedure was suggested by M09, and consists in determining whether the distance to the closest cell wall is greater than a few times the Rosseland mean free path:
\begin{equation}
d_{\rm min} > \frac{\gamma}{\rho\bar{\chi}_R}
\end{equation}
The $\gamma$ parameter controls the balance of speed versus accuracy. If $\gamma$ is set too low, then the sphere used for the MRW may be optically thin at long wavelengths, and therefore the photons might have escaped directly rather than diffuse if the MRW had not been used. This leads to over-estimated temperatures, as described in more detail in M09. If $\gamma$ is set too high, then the performance gains from using the MRW are decreased, since the photons still need to undergo a significant number of interactions. The starting criterion can be checked after each scattering or absorption/re-emission, and does not increase the computing time noticeably.

If the starting criterion is met, a sphere of radius $R_0$ and centered on the current photon position can be set up, with $R_0$ being at most the distance to the closest grid cell wall. The diffusion approximation can then be solved exactly inside the sphere (see M09 for details). The diffusion solution leads to the following algorithm: first, one samples a random number $\zeta\in[0,1]$ and uses it to solve for $y$ in the following equation:
\begin{equation}
\zeta = 2\,\sum_{n=1}^{\infty}\,(-1)^{n+1}\,y^{n^2}
\end{equation}
As $y$ tends to 1, the sum needs to be computed to higher and higher values of $n$ in order to preserve a constant numerical accuracy. The most efficient way to carry out this sampling is to pre-compute the sum very accurately for a range of y values, and to then interpolate for $y$ given $\zeta$ in the Monte-Carlo code. Once the value of $y$ is determined, one can compute the distance traveled to exit the diffusion sphere, using
\begin{equation}
ct = -\ln{y}\,\left(\frac{R_0}{\pi}\right)^2\,\frac{1}{D}
\end{equation}
where $D$ is the diffusion coefficient given by $D = 1/ 3\rho\bar{\chi}_R$. The energy absorbed by the dust grains (which is needed to calculate the temperature in both the \citealt{Lucy:99:282} and \citealt{Bjorkman:01:615} methods) during this MRW is then
\begin{equation}
E = E_\gamma\,ct\,\rho\,\bar{\kappa_P}
\end{equation}
where $E_ \gamma $ is the energy of the photon, $\rho$ is the density, and $\bar{\kappa_P}$ is the Planck mean mass absorption coefficient.

Finally, the emergent spectrum from an optically thick region is given by $I_\nu=S_\nu=B_\nu(T)$, so the frequency of the photons exiting the diffusion sphere should be sampled from the Planck function at the local dust temperature. If the \citet{Bjorkman:01:615} temperature correction method is used, the frequency of the photons should be sampled from $dB_\nu(T)/dT$ rather than simply $B_\nu(T)$.

\section{Summary}

In this note, I have shown that the MRW equations presented by M09 for the emission coefficient of dust, the diffusion coefficient, and the mean opacity to absorption can be simplified considerably. The emission coefficient defined by M09 does not need to be solved iteratively, but instead is directly given by Equation (\ref{eq:finaleta_complete}). The expression for the diffusion coefficient including scattering is in fact identical to that without scattering, and is given by Equation (\ref{eq:diffcoeff}). Finally, the mean opacity to absorption is simply the Planck mean mass absorption coefficient. All of these values are directly related to the Planck and Rosseland mean mass extinction and absorption coefficients, which are usually already pre-computed in Monte-Carlo radiation transfer codes. Thus, the number of calculations involved with the MRW can be greatly reduced, and the MRW technique can be implemented into existing codes with very little effort. An overview of the algorithm, including caveats, is given in \S\ref{sec:implementation}.

\begin{acknowledgements}
I wish to thank the referee for a careful review, and for insightful comments that improved this research note. I also wish to thank Barbara Whitney, Kenny Wood, and Katharine Johnston for useful discussions. Support for this work was provided by NASA through the Spitzer Space Telescope Fellowship Program, through a contract issued by the Jet Propulsion Laboratory, California Institute of Technology under a contract with NASA.
\end{acknowledgements}

\end{document}